\documentclass[epj]{svjour}

\usepackage{graphicx}
\usepackage{fancyhdr}
\def\makeheadbox{{
 }}
\def\beq{\begin{equation}}
\def\eeq#1{\label{#1}\end{equation}}
\def\eeqn{\end{equation}}
\def\beqa{\begin{eqnarray}}
\def\eeqa#1{\label{#1}\end{eqnarray}}
\def\eeqan{\end{eqnarray}}

\def\leqn#1{(\ref{#1})}

\def\mm{\tilde{m}_1}
\def\mp{\tilde{m}_2}

\def\tl{\tilde{t}_1}
\def\th{\tilde{t}_2}

\def\Mstop{M_{\tilde{t}}}

\newcommand{\bspace}{\!\!\!\!}
\def\met{\mbox{$E{\bspace}/_{T}$}}

\def\stacksymbols #1#2#3#4{\def\theguybelow{#2}
    \def\vp{\lower#3pt}
    \def\sp{\baselineskip0pt\lineskip#4pt}
    \mathrel{\mathpalette\intermediary#1}}

\def\intermediary#1#2{\vp\vbox{\sp
     \everycr={}\tabskip0pt
     \halign{$\mathsurround0pt#1\hfil##\hfil$\crcr#2\crcr
              \theguybelow\crcr}}}

\def\gsim{\stacksymbols{>}{\sim}{2.5}{.2}}
\def\lsim{\stacksymbols{<}{\sim}{2.5}{.2}}

\def\mytitle{My title} 
\def\myauthors{My name}  
\def\mytype{My type of session}
\def\mysession{My session}

\def\mytitle{The MSSM golden region and its collider signature} 
\def\myauthors{Maxim Perelstein, Christian Spethmann}  
\def\mytype{Contributed Talk}    
\def\mysession{Colliders - SUSY Phenomenology}

\begin{document}
\title{The MSSM golden region and its collider signature}
\author{Maxim Perelstein
 \and
 Christian Spethmann
}                     
%
%
\institute{\it Cornell Institute for High-Energy Phenomenology, 
Cornell University, Ithaca, NY~14853}
%
\date{}
\abstract{
The ``golden region'' in the Minimal Supersymmetric Model (MSSM)
parameter space is the region where the 
experimental constraints are satisfied and the amount of fine-tuning is 
minimized. In this region, the stop trilinear soft term $A_t$ is large,
leading to a significant mass splitting between the two stop mass 
eigenstates. As a result, the decay $\th\to\tl Z$ is kinematically
allowed throughout the golden region. We propose that the experiments 
at the Large Hadron Collider (LHC) can search for this decay through an 
inclusive signature, $Z+2j_b+\met+X$. We evaluate the Standard Model 
backgrounds for this channel, and identify a set of cuts that would 
allow detection of the supersymmetric contribution at the LHC for 
the MSSM parameters typical of the golden region. 
} 
\maketitle
%

\section{The MSSM Golden Region}

In this talk (based on our recent paper~\cite{PS}) we explore the idea that data and naturalness point to a particular "golden'' 
region within the Higgs and top sectors of the minimal 
supersymmetric model (MSSM), where the experimental 
bounds from non-observation of superpartners and the Higgs boson 
are satisfied and fine-tuning is close to the minimum value possible.  

We start by outlining the boundaries of the MSSM golden region. 
These are somewhat fuzzy due to an 
inherent lack of precision surrounding the concept of fine-tuning.
Our goal is to understand the qualitative features by 
making approximations which greatly clarify the picture.
 
The Higgs sector of the MSSM is strongly coupled to the top 
sector, but couplings to the rest of the MSSM are weaker. One may therefore 
begin by considering the Higgs and top sectors in isolation; 
that is, the gauge and non-top Yukawa couplings are set to zero. 
This is a good approximation as long as 
$
M_1/\Mstop\lsim 4,~~M_2/\Mstop \lsim 2,~~ 
M_3/\Mstop \lsim 10,
M_{\tilde{b}}\lsim 35 \Mstop / \tan\beta,$
where $g_{1,2,3}$ and $M_{1,2,3}$ are the gauge couplings and (weak-scale) gaugino masses 
for $U(1)$, $SU(2)$ and \linebreak $SU(3)$, and $\Mstop$ 
and $M_{\tilde{b}}$ are the stop and sbottom mass scales.
In this approximation, physics is described in 
terms of the holomorphic Higgs mass $\mu$ and the six parameters appearing 
in the soft Lagrangian for the Higgs and top sectors.
Since the model has to reproduce the known EWSB scale, $v=174$ GeV, 
only six parameters are independent. We choose the physical basis:
\beq
\tan\beta, \mu, m_A, \mm, \mp, \theta_t,
\eeq{parameters}
where $m_A$ is the CP-odd Higgs mass, $\mm$ and $\mp$ are stop eigenmasses
(by convention, $\mp>\mm$) and $\theta_t$ is the stop mixing 
angle. We will analyze the fine-tuning and Higgs mass constraints in 
this approximation and map out the golden region in the six-parameter 
space~\leqn{parameters}.

\begin{figure}[tb]
\begin{center}
\includegraphics[width=.35\textwidth]{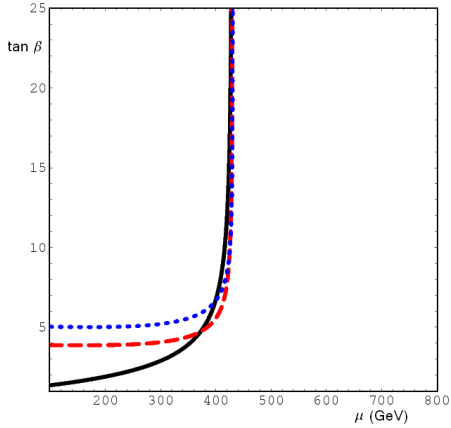}
\vskip2mm
\caption{Contours of 1\% fine-tuning in the ($\mu, \tan\beta$) plane.
The black (solid) contour corresponds to $m_A=100$ GeV, 
the red (dashed) and blue (dotted) contours correspond to $m_A=1.5$ 
and 2 TeV, respectively.}
\label{fig:FT}
\end{center}
\end{figure}

Following Barbieri and Guidice~\cite{BG}, we quantify fine-tuning by 
computing
\beq
A(\xi)\,=\,\left| \frac{\partial\log m_Z^2}{\partial\log \xi}\right|,
\eeq{ftpars}
where $m_Z$ is the tree-level $Z$ mass in the MSSM and 
$\xi=m_u^2, m_d^2, b, \mu$ are the relevant Lagrangian parameters.
The overall fine-tuning $\Delta$ is defined
by adding the four $A$'s in quadruture. 
For concreteness, we will require $\Delta\leq 100$, 
corresponding to fine tuning of 1\% or better. This requirement maps out
the golden region in the space of $(\tan\beta, \mu, M_A)$, as illustrated in
figure~\ref{fig:FT}. 

\begin{figure*}
\includegraphics[width=0.33 \textwidth]{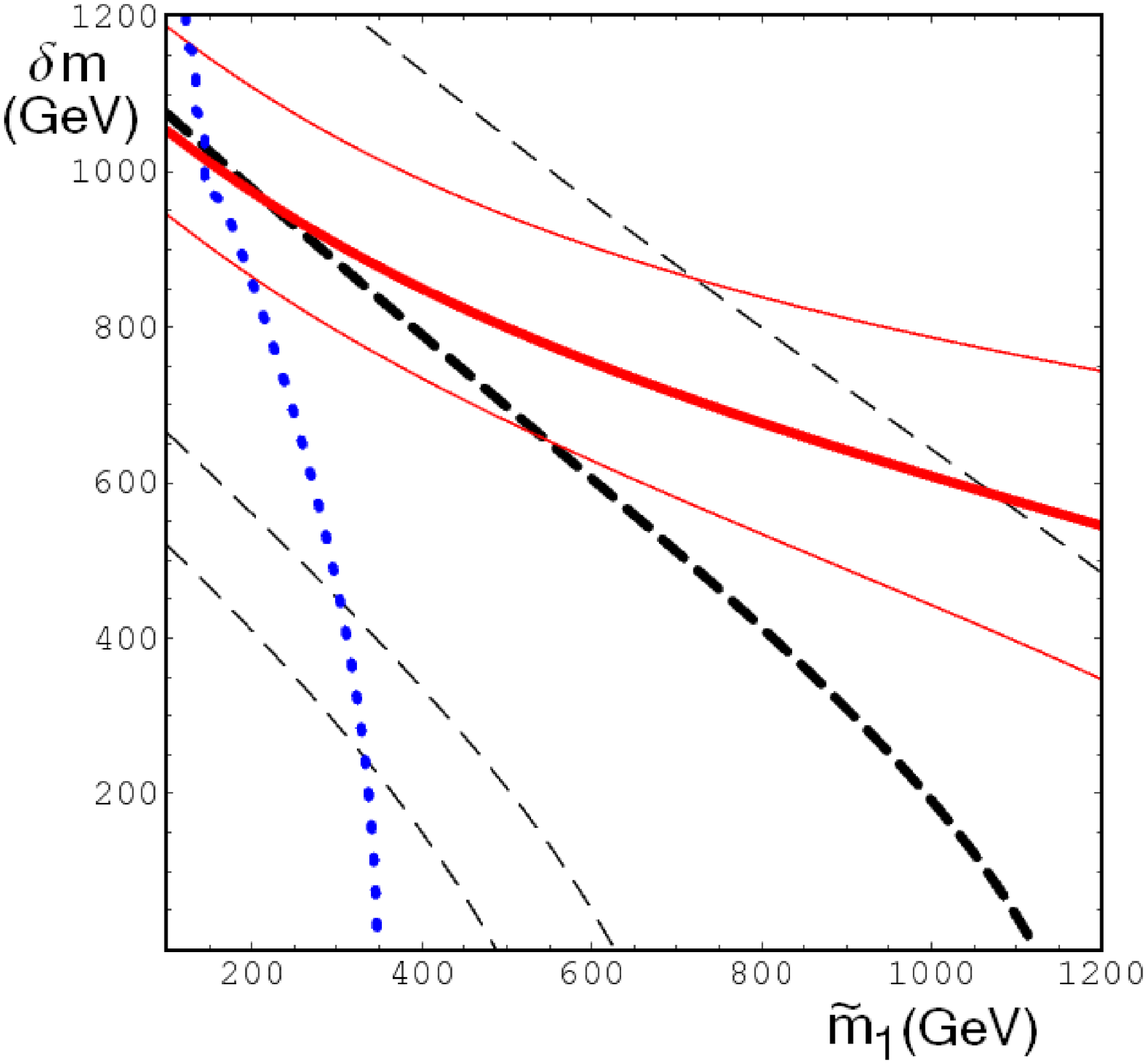}
\includegraphics[width=0.33 \textwidth]{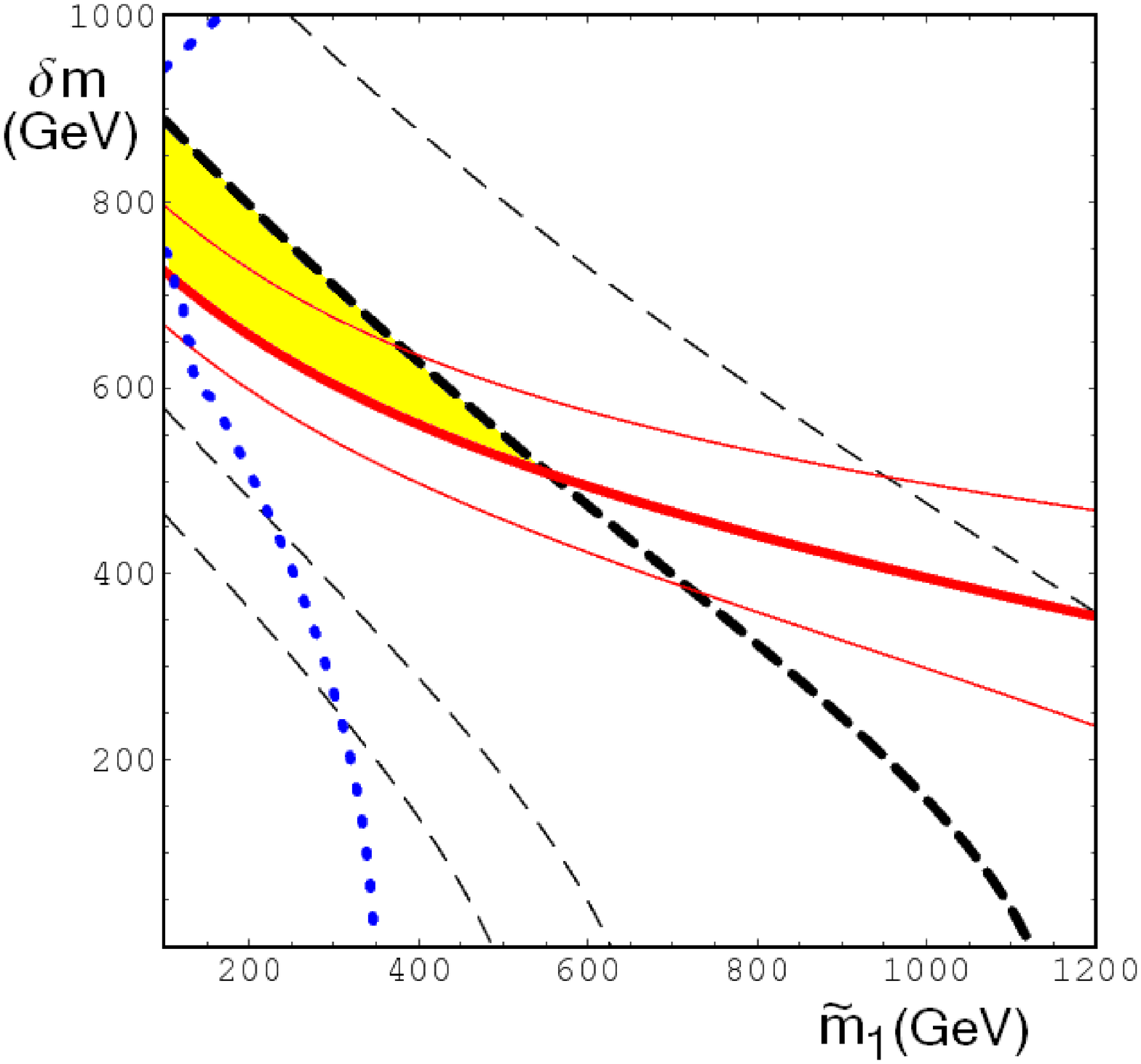}
\includegraphics[width=0.33 \textwidth]{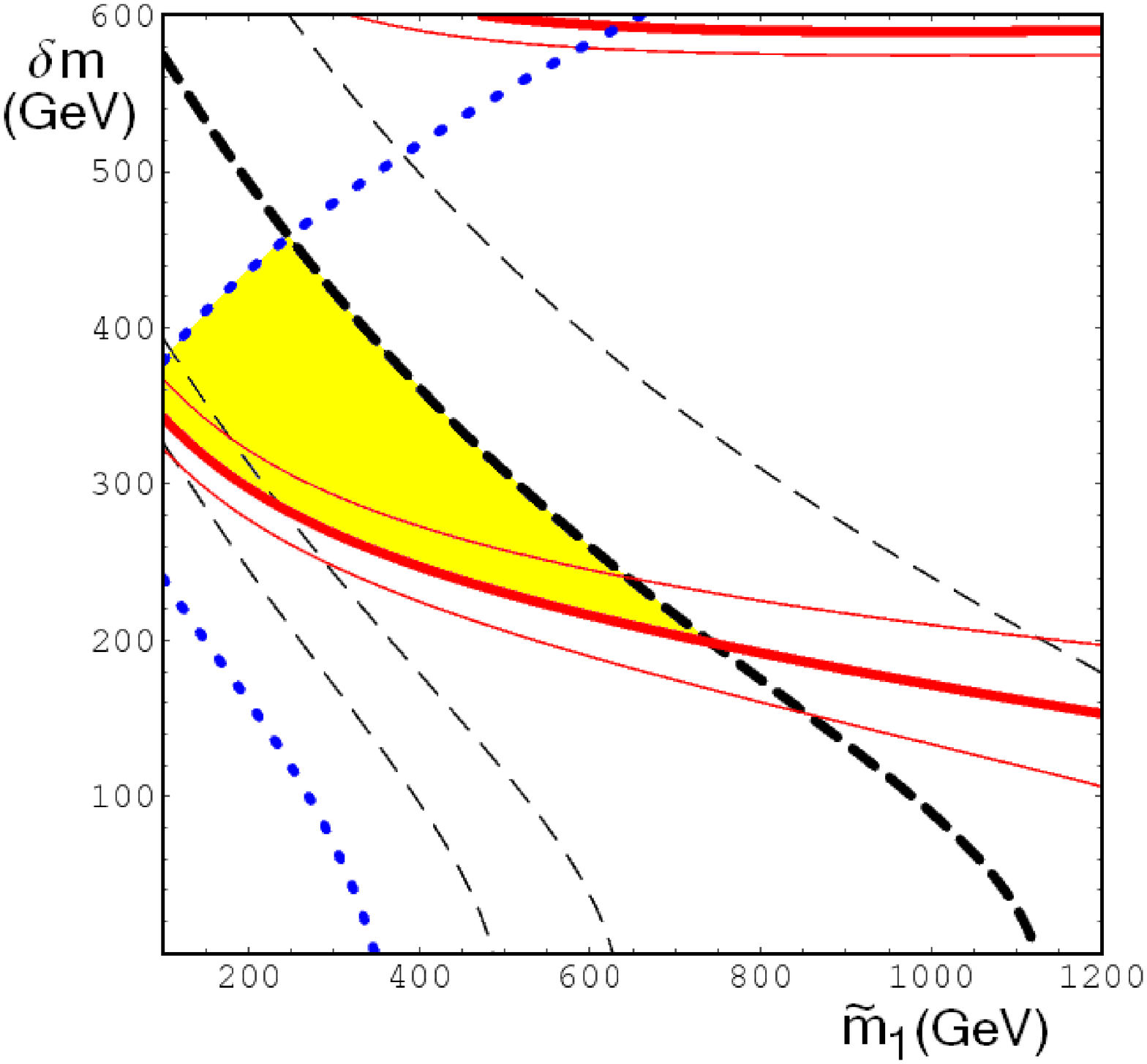}
\vskip2mm
\caption{Fine-tuning (black/dashed contours), Higgs mass bound
(red/solid contours), and $\rho$-parameter (blue/dotted contours)
constraints in the ($\mm$, $\delta m$) plane. The three panels 
correspond to: $\theta_t=\pi/25, \pi/15, \pi/4$. 
In all panels $\tan\beta=10$. The
yellow/shaded intersection of the regions allowed by the three constraints
is the MSSM ``golden'' region.}
\label{fig:FTloop}
\end{figure*}

The largest one-loop quantum correction to the $Z$ mass in the
MSSM is due to the contribution to $\delta m_{H_u}^2$ from top and stop 
loops: 
\beq
\delta_t m_Z^2 \approx -\delta m_{H_u}^2\left( 1-\frac{1}{\cos 2 \beta} 
\right),
\eeq{tloopZ}
where we ignored the renormalization of the angle $\beta$ by top/stop 
loops (the contribution of this effect scales as $1/\tan^2\beta$ and is 
subdominant for $\tan\beta\gsim2$). To measure the
fine-tuning between the bare (tree-level) and one-loop contributions, we
introduce
\beq
\Delta_t \,=\, \left|\frac{\delta_t m_Z^2}{m_Z^2}\right|. 
\eeq{deltat}
Choosing the maximum allowed value of $\Delta_t$ selects a region in the
stop sector parameter space, $(\mm, \mp, \theta_t)$, whose shape is 
approximately independent of the other parameters. This constraint is
shown by the black \linebreak (dashed) lines in Figs.~\ref{fig:FTloop}, where we plot 
5\%, 3\%, 1\% and 0.5\%
tuning contours (corresponding to $\Delta_t=20, 33.3, 100$, and 200,
respectively) in the stop mass plane for several values of $\theta_t$
and $\tan\beta=10$. 
 
The second constraint that determines the shape of the golden region is the 
LEP2 lower bound on the Higgs mass~\cite{LEPHiggs}. For generic MSSM parameter
values, the limit on the lightest CP-even Higgs is very close to that for
the SM Higgs, $m(h^0)\gsim 114~{\rm GeV}$, 
and large loop corrections are required to satisfy this bound. We used a 
simple analytic approximation, due to 
Carena {\it et.~al.}~\cite{MHanal}, which includes the one-loop and 
leading-log two-loop contributions from top and stop loops.
It agrees with the state-of-the-art calculations to within a
few GeV for typical MSSM parameters~\cite{MHfull}.

The contours in the stop mass plane corresponding to the LEP2 Higgs mass 
bound are superimposed on the fine-tuning contours in Figs.~\ref{fig:FTloop}.
The positions of these contours depend strongly on the top quark mass. We used
$M_t=171.4\pm 2.1$ GeV~\cite{mtop}, and plotted the constraint 
corresponding to the central value (thick red/solid lines),
as well as the boundaries of the 95\% c.l. band (thinner red/solid lines). 
The contours are 
approximately independent of $\tan\beta$ for $3\lsim\tan\beta\lsim 35$; 
the golden region shrinks rapidly outside of this range of $\tan\beta$.
We use $\tan\beta=10$ in the plots. 
The overlap between the regions of acceptably low fine-tuning (for
definiteness, we choose $\Delta_t=100$) and experimentally allowed Higgs mass 
defines the golden region, shaded in yellow in Figs.~\ref{fig:FTloop}.

LEP2 searches for direct production of charginos and 
stops constrain both $\mu$ and $\mm$ to be above $\approx 100$ GeV, and are 
largely independent of the rest of the MSSM parameters. 
The Tevatron stop searches yield a similar (though more
model-dependent) bound on $\mm$. 

In the presence of a large $A_t$ term, stop and sbottom loops 
may induce a significant correction to the $\rho$ parameter. 
For our purposes, it suffices to use the one-loop result. 
Using the PDG value $\rho = 1.0002  \begin{array}{l} \scriptstyle +0.0004 
\\[-1.3ex] \scriptstyle -0.0007 \end{array}$~\cite{PDG}, we obtain the 
95\% c.l. contours in the stop mass plane shown by the blue/dotted lines in 
Figs.~\ref{fig:FTloop}. This constraint eliminates a part of the parameter 
space with very low $\mm$ and large $\delta m$. 

The supersymmetric contribution to $g_\mu-2$ 
depends sensitively on the slepton and weak gaugino mass scales, and only 
weakly on the parameters defining the golden region. 
We can expect a large contribution to the $b\to s\gamma$ rate 
from the $\tilde{t}-\tilde{H}$ loop. 
It is well known, however, that this can be cancelled
by the contribution of the top-charged Higgs loop. A simplified analysis
of this constraint based on ~\cite{bsgamma} shows that for 
{\it any} values of the stop masses 
inside the golden region, and for any value of
$\mu$ between 100 and 500 GeV, one can find values of $m_A$ in the 
100-1000 GeV range for which this cancellation ensures consistency with 
experiment. 

\section{The expected LHC signature}
\label{obs}

The golden region has the following interesting
qualitative features:
\begin{itemize}
\item Both stops typically have masses below $1$ TeV; 
\item A substantial mass splitting between the two stops:
typically, $\delta m\gsim 200$ GeV; 
\item The stop mixing angle must be non-zero
\end{itemize}
The first feature implies that both $\tl$ and $\th$ will be produced with
sizeable cross sections at the LHC. 
The second feature implies that the decay mode
$\th\to \tl Z$ is kinematically allowed. 
The last feature guarantees that the vertex is 
non-zero, and the decay $\th\to \tl Z$ indeed occurs.

The branching ratio of the $\th\to \tl Z$ mode depends on which competing 
$\th$ decay channels are available. The possible two-body channels 
are
\beq
t \tilde{g},~~t \tilde{\chi^0},~~b \tilde{\chi^+}\,,
~~\tilde{b} W^+,~~\tilde{b} H^+,~~\tilde{t}_1 h^0,~~
\tilde{t}_1 H^0,~~\tilde{t}_1 A^0\,,
\eeq{chan}
where $\tilde{\chi^0}$ and $\tilde{\chi^+}$ denote all neutralinos and
charginos that are kinematically accessible, and flavor-changing couplings 
are assumed to be negligible.  

For a more detailed analysis, we choose a benchmark point (BP)
representative of the golden region (see Table~\ref{tab:bpdef}). 
The physical stop masses are $\mm = 400$ GeV and $\mp = 700$ GeV, with
maximal mixing $\theta_t=\pi/4$.
Using {{\tt SuSpect}\cite{suspect}, we checked that the Higgs mass, the 
$b\to s\gamma$ branching ratio, the $\rho$ parameter and the supersymmetric 
contribution to the muon anomalous magnetic moment at the BP are consistent 
with the current experimental constraints. 
The $\th\to\tl Z$ mode has a substantial branching ratio, about 31\%. 

\begin{table*}[t!]
\caption{The benchmark point: MSSM parameters, defined at the weak scale
(dimensionful parameters in GeV).}
\begin{tabular}{lllllllllllll} \hline\noalign{\smallskip}
$m_{Q^3}$ & $m_{u^3}$ & $m_{d^3}$ & $A_t$ & $\mu$ & $m_A$ & $\tan\beta$ 
& $M_1$ & $M_2$ & $M_3$ & $m_{\tilde{q}}$ & $m_{\tilde{\ell}}$ \\ 
\noalign{\smallskip}\hline\noalign{\smallskip}
548.7 & 547.3 & 1000 & 1019 & 250 & 200 & 10 & 1000 & 1000 & 1000 & 
1000 & 1000 \\ \hline
\end{tabular}
\label{tab:bpdef}
\end{table*}

At the benchmark point, we find 
$\sigma(pp\to \th \th^*)= 0.05$ pb.
In about 52\% of the events, either one or both 
of the produced stops decays in the $\tl Z$ mode. This decay is followed by 
a cascade 
\beqa
\tl \to \chi_1^+ b,~~~\chi_1^+ &\to& u \bar{d} \chi_1^0~/~c \bar{s} \chi_1^0
                             ~/~ \ell^+ \nu \chi_1^0, 
\eeqa{cascade}
where the jets and leptons are 
very soft due to a small chargino-neutralino mass splitting.  

Both the $\tl$ and $\th$ decays always 
produce a $b$ quark, as well as (assuming conserved R parity 
and a weakly interacting lightest 
supersymmetric particle) large missing 
transverse energy.
In order to make the analysis as model-independent as possible, we
therefore focus on an inclusive signature, 
\beq
Z(\ell^+,\ell^-) + 2 j_b + \met + X,
\eeq{sign}
where $Z(\ell^+,\ell^-)$ denotes a lepton pair ($\ell=e$ or $\mu$) with
the invariant mass at the $Z$ peak. 
The presence of energetic leptons ensures that essentially all such events 
will be triggered on. 

To assess the observability of the signature~\leqn{sign}, 
we have simulated statistically significant event samples for the signal
and the most relevant SM backgrounds using the 
{\tt MadGraph/} {\tt MadEvent v4} software package~\cite{MG},
which includes {\tt Pythia}~\cite{Pythia} and PGS.

The irreducible 
SM backgrounds that were analyzed in detail are $jjZZ$, $t \bar{t}Z$ 
and $t \bar{t}$, all of which can genuinely produce the signature~\leqn{sign}.
To isolate the signal events, we imposed the following set of cuts: 

\begin{enumerate}

\item Two opposite-charge same-flavor leptons with\\
$\sqrt{s(\ell^+\ell^-)}=M_Z\pm2$ GeV.

\item Two hard jets with $p_T>125$ GeV for the first and
$p_T>50$ GeV for the second;

\item At least one of the two highest-$p_T$ jets must be $b$-tagged;

\item The boost factor of the reconstructed $Z$ boson, $\gamma(Z)=1/\sqrt{1-v_Z^2}$,  
must be larger than 2.0;

\item A missing $E_T$ cut: $\met>225$ GeV.

\end{enumerate}

\begin{table*}[t!]
\caption{Observability of the
golden region signature~\leqn{sign}. First row: Production cross sections 
at the LHC. Second row: Number of 
Monte Carlo events. Rows 3--8: Cut efficiencies, in\%. 
Last row: Expected number of events for 100 fb$^{-1}$.}
\begin{tabular}{llllll} \hline\noalign{\smallskip}
& signal: $\tilde{t}_2\tilde{t}_2^*$ & $jjZZ$& $t\bar{t}Z$& $t\bar{t}$& $jjZ$ 
\rule{0ex}{2.2ex} \\ 
\noalign{\smallskip}\hline\noalign{\smallskip}
$\sigma_{\rm prod}$(pb) & 0.051& 0.888& 0.616& 552& 824\\ 
total simulated & 9964& 159672& 119395& 3745930& 1397940\\ 
\noalign{\smallskip}\hline\noalign{\smallskip}
1. leptonic $Z$(s) & 1.4& 4.5& 2.6& 0.04& 2.1 \\ 
2(a). $p_t(j_1) >125$ GeV & 89& 67& 55& 21& 41\\ 
2(b). $p_t(j_2) >50$ GeV & 94& 93& 92& 76& 84\\ 
3. $b$-tag & 64 & 8& 44& 57& 5\\ 
4. $\gamma(Z)>2.0$ & 89 & 66& 69& 26& 68\\ 
5. $\met > 225$ GeV & 48& 2.2& 4.4& 1.7& $<0.9$ (95\% c.l.); 0 (ext.)\\ 
\noalign{\smallskip}\hline\noalign{\smallskip}
$N_{\rm exp}$(100 $fb^{-1}$) & 16.4 & 2.8 & 10.8 & 8.8 &  $<177$ (95\% c.l.); 0 (ext.)\\ 
\noalign{\smallskip}\hline\noalign{\smallskip}
\end{tabular}
\label{tab:cuts}
\end{table*}

\noindent  The background production cross sections at the LHC and the
efficiencies of these cuts are given in Table~\ref{tab:cuts}.

Assuming that the search is statistics-limited, 
we estimate that a 3-sigma observation
would require 75 fb$^{-1}$ of data, while a definitive 5-sigma discovery is
possible with 210 fb$^{-1}$. Note that the $t\bar{t}$ background can
be effectively measured from data by measuring the event rates with 
dilepton invariant masses away from the $Z$ peak and performing shoulder 
subtraction. 

We also considered several other irreducible SM backgrounds which
are expected to be less significant, 
but might nevertheless be relevant. The most 
important one is $t\bar{t}j$, where $j$ is a hard jet. The cross 
section for this channel is suppressed compared to $t\bar{t}$, but the 
presence of the additional hard jet increases the probability that the events 
will pass the jet $p_T$ cut (cut 2). We find a parton-level cross 
section $\sigma(t\bar{t}j, p_{T}^j>125~{\rm GeV})=65$ pb. Assuming 
conservatively that all these
events pass cut 2 and that the efficiencies of all other cuts are the 
same as for the $t\bar{t}$ sample, we expect that this background would 
add at most about 50\% to the $t\bar{t}$ rate. As in the $t\bar{t}$ case, 
this contribution can be subtracted using data away from the $Z$ peak in the
lepton invariant mass distribution. Assuming that the statistical error 
dominates
this subtraction, the net effect would be an increase in the integrated 
luminosity required to achieve the same level of significance by at most
about 10\%.

Other backgrounds we considered are three vector boson channels
$ZZZ$, $ZZW$, and $ZWW$; as well as channels with single top production,  
$tZj$ and $\bar{t}Zj$. Combining the parton-level cross sections for these 
channels with the branching ratios of decays producing the 
signature~\leqn{sign} results in event rates that are too small to affect 
the search. 

While the SM processes considered above genuinely produce the 
signature~\leqn{sign}, other SM processes may contribute to the background
due to detector imperfections. We expect that the dominant among these
is the process $jjZ$, with $Z\to \ell^+\ell^-$ and apparent $\met$ 
due to jet energy mismeasurement or other instrumental issues. We
conducted a preliminary investigation of this background by generating 
and analyzing a sample of $1.4\times 10^6$ $jjZ$ events with
$p_{T, {\rm jet}}^{\rm min}=50$ GeV (see the last column 
of Table~\ref{tab:cuts}). None of the events in this sample pass the
cuts 1-5. This allows us to put a 95\% c.l. bound on the combined 
efficiency of
this set of cuts for the $jjZ$ sample of about $2\times 10^{-6}$, 
corresponding to
a background rate about 10 times larger than the signal rate. 
However, we expect that the actual $jjZ$ background rate is well below 
this bound, since all 349 events in our sample that pass the cuts 1--4 
in fact have $\met$ below 50 GeV. We find that the $\met$ distribution
of these 349 events can be fit with an exponential, $N\propto 
e^{-0.10\met}$, where $\met$ is in units of GeV. Assuming that this scaling 
adequately describes the tail of the distribution at large $\met$, we
estimate that the rate of $jjZ$ events passing all 5 cuts is 
completely negligible and that this background
should not present a problem. 

This conclusion is
of course rather preliminary, and this issue should be revisited 
once the performance of the LHC detectors is understood using real data.
Note that the necessity to understand the shape and normalization of the 
large apparent $\met$ tail from SM processes with large cross sections is 
not unique to the signature discussed here, but is in fact crucial for most 
SUSY searches at the LHC.

\section{Conclusions}

Our analysis indicates that at the BP 
the signature~\leqn{sign} can be discovered at the LHC. The chosen BP is typical of
the golden region, and this conclusion should generally hold as the 
MSSM parameters are varied away from the BP. There are,
however, several exceptional parts of the parameter space where the
observability of this signature could be substantially degraded:

\begin{itemize}
\item Large $\mp$ region: The $\th$ production cross section drops rapidly 
with its mass;
\item Small $\theta_t$ region: 
The branching 
ratio Br($\th\to Z\tl$) is proportional to $\sin^2 2\theta_t$;
\item Small $\tl$-LSP mass difference: The absence of hard jets in this case 
would make the signal/background discrimination more difficult.
\end{itemize}

Unfortunately, a positive identification of non-SM physics in 
this channel would {\it not} necessarily imply that the stops are split,
since other MSSM processes may produce the same signature: for example,
$Z$'s can be produced in neutralino or chargino cascade decays.
Careful comparisons of the rates with and without $b$ jets, 
as well as the distribution of events in vector boson-jet 
invariant masses, can remove the ambiguity as explained in \cite{PS}. 
This may however take considerably more data
than the discovery of an excess over the SM backgrounds in these channels.

Given the strong theoretical motivation for the signature discussed here,
we encourage experimental collaborations to perform a more detailed 
study of its observability. If the first round of the LHC results 
points towards an MSSM-like theory, obtaining experimental 
information about the stop spectrum, and in particular
testing whether the ``golden region MSSM'' hypothesis is correct, will
become an important priority for the LHC experiments.


\begin{thebibliography}{99}

\bibitem{PS}
  M.~Perelstein and C.~Spethmann,
  JHEP {\bf 0704}, 070 (2007)
  [arXiv:hep-ph/0702038].
  
\bibitem{BG}
  R.~Barbieri and G.~F.~Giudice,
  Nucl.\ Phys.\ B {\bf 306}, 63 (1988).

\bibitem{LEPHiggs}
  S.~Schael {\it et al.}  [ALEPH Collaboration],
  Eur.\ Phys.\ J.\ C {\bf 47}, 547 (2006)

\bibitem{PDG}
  W.~M.~Yao {\it et al.}  [Particle Data Group],
  J.\ Phys.\ G {\bf 33}, 1 (2006).

\bibitem{MHfull}
  G.~Degrassi, S.~Heinemeyer, W.~Hollik, P.~Slavich and G.~Weiglein,
  Eur.\ Phys.\ J.\ C {\bf 28}, 133 (2003)

\bibitem{suspect}
  A.~Djouadi, J.~L.~Kneur and G.~Moultaka,
  arXiv:hep-ph/0211331.

\bibitem{MHanal}
  M.~Carena, J.~R.~Espinosa, M.~Quiros and C.~E.~M.~Wagner,
  Phys.\ Lett.\ B {\bf 355}, 209 (1995)

\bibitem{mtop}
  E.~Brubaker {\it et al.}  [Tevatron Electroweak Working Group],
  arXiv:hep-ex/0608032.

\bibitem{bsgamma}
  R.~Barbieri and G.~F.~Giudice,
  Phys.\ Lett.\ B {\bf 309}, 86 (1993)

\bibitem{MG}
  F.~Maltoni and T.~Stelzer,
  JHEP {\bf 0302}, 027 (2003).

\bibitem{Pythia}
  T.~Sj\"ostrand {\sl et al.},
  Comput.\ Phys.\ Commun.\  {\bf 135}, 238 (2001);
  T.~Sjostrand, S.~Mrenna and P.~Skands,
  JHEP {\bf 0605}, 026 (2006)

\bibitem{WY}
  L.~T.~Wang and I.~Yavin,
  arXiv:hep-ph/0605296.

\end{thebibliography}
\end{document}